\documentclass[amssymb,amsfonts,aps,prb,twocolumn,longbibliography]{revtex4-1}

\usepackage[utf8]{inputenc}

\usepackage{graphicx}
\usepackage{dcolumn}
\usepackage{bm}
\usepackage[normalem]{ulem}
\usepackage{xcolor}
\usepackage{comment}
\usepackage{amsmath}
\usepackage{siunitx}
\usepackage{dsfont}
\usepackage{float}
\usepackage{placeins}
\usepackage[shortlabels]{enumitem}
\usepackage{soul}
\newcommand{\redmark}[1] {\color{red}#1\color{black}}

\DeclareSIUnit\angstrom{\text{Å}}

\usepackage{physics}
\begin{document}

\title{
Characterizing $S=3/2$ AKLT Hamiltonian with Scanning Tunneling Spectroscopy}

\author{
M. Ferri-Cort\'es$^{1}$, 
J. C. G. Henriques$^{2,3}$, 
J. Fern\'andez-Rossier$^{3}$
}
\altaffiliation{
On permanent leave from Departamento de F\'isica Aplicada, Universidad de Alicante, 03690 San Vicente del Raspeig, Spain.
}

\affiliation{$^1$ Departamento de F\'isica Aplicada, Universidad de Alicante, 03690 San Vicente del Raspeig, Spain}

\affiliation{
$^2$ Universidade de Santiago de Compostela, 15782 Santiago de Compostela, Spain}

\affiliation{
$^3$International Iberian Nanotechnology Laboratory (INL), Av. Mestre Jos\'e Veiga, 4715-330 Braga, Portugal
}

\date{\today}

%------------------------------------------------------------------------
\begin{abstract} 
The AKLT Hamiltonian is a particular instance of a general class of
model Hamiltonians defined in lattices with coordination $z$ where each site
hosts a  spins $S=z/2$, interacting both with linear and non-linear exchange couplings.
In two dimensions, the AKLT  model features a gap in the spectrum, and  its ground state is a valence bond solid state;
that is an  universal resource for measurement based quantum computing, motivating the quest of physical systems that realize this Hamiltonian.
Given a   finite-size system described with a specific instance of this general class of models, we address the question of how
to asses if such system is a realization of the AKLT model using inelastic tunnel spectroscopy implemented with scanning tunnel microscopy (IETS-STM).
We propose two approaches. First,  in the case of a dimer, we show how to leverage non-equilibrium IETS-STM to obtain the energies of all  excited states, 
and determine thereby the magnitude of both linear and non-linear exchange interactions.  Second, we explore how IETS can probe the in-gap excitations associated to edge spins. In the AKLT limit, spins $S=3/2$ at the edge of the lattice have  coordination 2, giving rise to $S=1/2$ dangling spins that can be probed with IETS.  We propose  a $S=1/2$ effective Hamiltonian to describe the interactions between these dangling spins in the neighborhood of the AKLT point, where their degeneracy lifted.
%Specifically, we find that above-gap excitations accessible with IETS correspond to  magnon-polaron states, where the $S=1/2$ dangling spins are fully polarized. \redmark{(Joao: We never discuss this magnon-polaron part)}

\end{abstract}

\maketitle

%------------------------------------------------------------------------
\section{Introduction}

Quantum computing hinges on certain quantum states that make it possible  to  solve a variety of important  problems\cite{dalzell23} outperforming classical hardware.  There are two complementary strategies for the generation of these resourceful  quantum states. In the gate-based quantum computing approach, quantum algorithms are a prescribed  sequence, most often very long, of quantum gates acting on one and two qubits at a time, starting from an initial unentangled product state. In this strategy, the system is driven out of equilibrium. 
In measurement-based quantum computing\cite{raussendorf03} (MBQC), the initial state is entangled in a special way, that makes it possible to implement quantum algorithms combining single-qubit gates and readouts, without having to use two-qubit gates.  
Importantly, the ground state of some Hamiltonians, such as the  AKLT model in two dimensions \cite{Affleck1987,affleck88}, belong to this class of initial entangled states and therefore said to be universal resources\cite{Wei2011} for MBQC.
In addition,  it was demonstrated that the AKLT model is gapped in two dimensions\cite{pomata20,lem20}.   Therefore, if it were possible to engineer a physical system so that it realizes the two-dimensional AKLT model, and at a temperature significantly smaller than the gap, such system would spontaneously provide a very good starting point for MBQC.

AKLT models\cite{Affleck1987,affleck88,wei22} are particular instances of a general class of
 model Hamiltonians describing  spins $S=z/2$ interacting both with linear and non-linear exchange couplings,  in lattices with coordination number $z$. The wave function of the ground state can be written explicitly as a valence bond solid state: in every physical site, the spin $S$ is represented by means of $z$ virtual spins $S=1/2$. At each bond between two physical sites, a singlet is formed between two virtual spins. Furthermore, the wave function is symmetrized at each physical site, to ensure that the $z$ virtual spins  $1/2$ realize a physical spin $S$.  In systems with periodic boundary conditions, the ground state is unique. In contrast, in systems with boundaries, the edge physical spins have dangling virtual spins, giving rise to a degeneracy of the ground state that can be interpreted in terms of emergent fractional $S=1/2$ degrees of freedom. This constitutes a canonical example of fractionalization.
 
 For spin chains  \cite{Affleck1987},  the AKLT model is given by the bilinear-biquadratic (BLBQ) Hamiltonian with $S= 1$: 
 \begin{equation}
     H_{BLBQ}= J \left( \sum_i\vec{S}_i\cdot\vec{S}_{i+1} +\beta (\vec{S}_i\cdot\vec{S}_{i+1})^2\right) 
     \label{eq:BLBQ}
 \end{equation}
 with $\beta=\frac{1}{3}$.  
 The most important  properties of the one-dimensional AKLT model are the gap in the excitation spectrum, the symmetry protected topological order\cite{Pollmann2012} and  the existence of fractional $S=1/2$ edge excitations in open-end chains,   shared\cite{schollwock96} by  the BLBQ family in all points between $\beta=0$, the Heisenberg model, and the AKLT point $\beta=1/3$.  These  properties define the Haldane phase and have been observed experimentally in a number of systems, 
 including crystals with decoupled spin chains\cite{Hagiwara1990}, cold-atoms\cite{sompet22} and, important for this work, artificial one-dimensional $S=1$ triangulene  lattices\cite{Mishra2021}. 
 Whereas these systems provide a unique arena to explore fractionalization and to test the Haldane prediction\cite{Haldane83},  the ground state of the Haldane phase of the BLBQ model  
  is a universal resource to implement 1-qubit gates only\cite{Wei2011} and, therefore, they  do not provide a viable alternative to gate-based quantum computing.

The simplest version of the AKLT 
model in 2D is realized in a honeycomb lattice\cite{affleck88}, where $z=3$ and $S=3/2$. The AKLT is a special instance of the  bilinear-biquadratic-bicubic (BLBCBQ) Hamiltonian:
 \begin{equation}
     H_{BLBQBC}= J_1 \left( \sum_{\langle i, i' \rangle} \sum_{n=1}^{3} \beta_n \left(\vec{S}_i\cdot\vec{S}_{i'}\right)^n\right) 
     \label{eq:BLBQBCHamiltonian}
 \end{equation}
 where $i$ runs over the sites of the lattice, $i'$ runs over the first neighbours of $i$ and
 $J_1$ is the magnitude of the  linear exchange (we consider $\beta_1=1$). The relative magnitude of non-linear exchange couplings is controlled by the parameters $\beta_2,\beta_3$. The AKLT point\cite{affleck88}, for which the AKLT state is the ground state of Hamiltonian (\ref{eq:BLBQBCHamiltonian}),  is reached when  $\beta_2=\frac{116}{243}, \beta_3=\frac{16}{243}$. We refer to the region in the $\beta_2,\beta_3$ plane that contains the AKLT point and has the same key properties ( $S=0$ ground state, gap in the excitation spectrum, fractional edge spins), as the AKLT phase. 
In contrast with the 1D case,  very few papers have explored theoretically the boundaries of the AKLT phase in 2D  \cite{Ganesh2011,garcia13,huang13}.
Also in contrast with the one-dimensional case, so far no physical system has been found that realizes the AKLT phase  in two dimensions. Promising physical platforms to make this happen would be multi-orbital Mott insulators \cite{sela15} and, given the successful precedent of $S=1$ Haldane chains,  lattices of $S=3/2$ nanographenes\cite{catarina2023,henriques2024}. 

The present work is inspired by the experimental approach that has been successful in the exploration of the 1D  Haldane phase both in artificial chains
made of $S=1$ nanographenes\cite{Mishra2021} as well as in the dimerized $S=1/2$ chains\cite{zhao2024a}. In these works, the use of inelastic electron tunnel spectroscopy (IETS) carried out with scanning tunnel microscope (STM), 
was instrumental to measure the Haldane gap and the in-gap excitations associated to fractional spins at the edges, as anticipated theoretically\cite{delgado13} and,  in the $S=1$ case\cite{Mishra2021}, to determine the presence of a significant non-linear exchange, $\beta\simeq 0.09$, that has been later accounted for theoretically\cite{henriques23}.   The fabrication of  small two-dimensional lattices of $S=3/2$ nanographenes\cite{delgado23}, for which finite non-linear exchange parameters have been computed\cite{catarina2023} is a step toward the realization of a physical system that realizes the two-dimensional AKLT phase.

Here we  address the question of how to certify, using STM-IETS, if a given on-surface spin system, including in particular a $S=3/2$ nanographene crystal %
 %molecular crystal 
 provides the physical realization of the AKLT phase.  Our work can be framed in the more general context of how to probe neutral excitations in quantum insulators\cite{wu24}.
 We focus on two different problems. First, how to infer the parameters of the BLBQBC Hamiltonian using STM-IETS
in a dimer. Second, we model the IETS-STM of the in-gap edge spin excitations in a hexamer of $S=3/2$ described by the  BLBQBC model. These in-gap states are expected in the AKLT point and its neighborhood and are akin to the extensively studied edge states in Haldane spin chains.
Our results stress the advantages of studying small lattices  to infer the spin couplings, as well as using the in-gap edge states as a smoking gun for the AKLT phase.  The feasibility of this  approach is backed up by   the experimental capability to fabricate and probe small artificial spin lattices, such as 
spin  dimers\cite{mishra2020, turco23} and hexamers \cite{Mishra2021,hieulle2021}, as well as other structures like $S=1/2$ spin chains\cite{zhao2024a,zhao2024b}. This bottom-up approach could be the way to avoid the trouble identifying the AKLT phase in macroscopically large systems,  a problem well illustrated in the field of Kitaev materials\cite{trebst22}.

The rest of this paper is organized as follows. In section II we review our methods. In section III we show the non-equilibrium IETS of a $S=3/2$ dimer described with the BLBQBC Hamiltonian. Next, in section IV we explore the  equilibrium IETS of an hexamer of $S=3/2$ described with the BLBQBC model,  both at the AKLT point and in the neighborhood. At last, in section V we present our final remarks.

\section{Methods }
In this work we will be considering both dimers and  hexamers of spins $3/2$, both because  triangulene dimers and hexamers with $S=1$ \cite{mishra2020,Mishra2021,hieulle2021} and with $S=1/2$ \cite{krane23,zhao2024b} have been reported, and also because the small size of these systems allows for an exact numerical diagonalization (ED)  of the BLBQBC Hamiltonian. In order to study the role of edge states in the AKLT phase, we  consider several  types of \emph{boundary} conditions, that differ by the number of \emph{passivated} edges of the hexamer. In the AKLT point, every spin with coordination $(3-n)$ has $n$ \emph{dangling} spin $S=1/2$ that do not form a singlet. 
\begin{figure}
    \centering
    \includegraphics[width=0.5\textwidth]{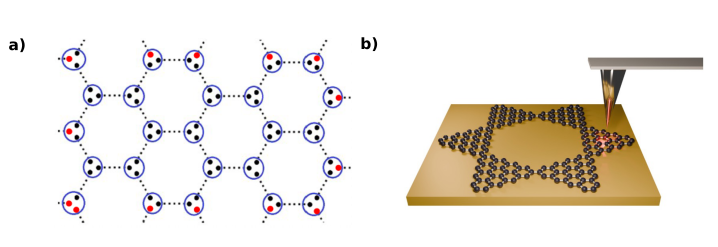}
    \caption{a) Honeycomb lattice with $S=3/2$. The edge states are shown in red. b) Schematic representation of a STM-IETS in a small cluster of $N=6$.}
    \label{fig:structure_BC}
\end{figure}

Reference values for $\beta_2,\beta_3$ which are relevant in this work are shown in the table. Specifically, we consider two $S=3/2$ molecules, the 4-triangulene (4T) \cite{catarina2023}, a triangular shape nanographene molecules with 4 benzene rings along its side,  and the 5-aza triangulene (5AT)
a triangulene with 5 benzene rings along its side, where the central carbon atom was replaced by a nitrogen one \cite{anindya2022controlling,pawlak2025surface,villasvarela2023surface,wang2022aza,lawrence2023topological,yu2024prediction}. Our calculations of $\beta_2,\beta_3$ for 4T\cite{catarina2023} the 5AT (see Appendix \ref{app:a})  use the same approach\cite{Mishra2021} that successfully accounts for non-linear exchange for $S=1$, namely, exact diagonalization of the Hubbard models in a restricted space of multielectronic configurations, and the mapping of the low energy levels to a spin Hamiltonian. 
\begin{table}[h!]
\centering
\begin{tabular}{|c|c|c|c|}
\hline
\textbf{System} & $\beta_2$ & $\beta_3$ & $J_1$ (meV) \\ \hline
AKLT model\cite{affleck88}  & $\frac{116}{243}$ & $\frac{16}{243}$ &  \\ \hline
$S=3/2$ triangulenes \cite{catarina2023} & 0.09 & 0.007 & 11.3 \\ \hline
5-aza triangulene (see appendix) & 0.27 & 0.034 & 10.6 \\ \hline
\end{tabular}
\caption{Relevant reference values for $\beta_2$ and $\beta_3$ considered in this work.}
\label{tab:beta_values}
\end{table}

Inelastic electron scanning tunneling microscopy can probe the energy excitations of a single surface spin\cite{heinrich04} through spin-flip assisted tunneling\cite{Fernandez09,lorente09,fransson09}, where transport electrons exchange spin and energy with the atomic spin (Fig. \ref{fig:structure_BC} b).
As electrons tunnel from tip to surface (or vice-versa), they can release their excess energy $eV$ to excite the on-surface spins.
The process has to conserve energy, so that $|eV| =E_M-E_{M'}$,  where $E_M,E_{M'}$ are  the energies of eigenstates of the on-surface spin Hamiltonian, the BLBQBC model in our case.   Hence, as the bias is ramped up, with either sign, new transport inelastic channels open, which results in step-wise increase of the conductance, $dI/dV$. Since the total spin has also to be conserved, and the tunneling electron can either conserve its spin or flip, we have $|S_M-S_{M'}|=0,1$. 

We compute the $dI/dV$ treating the Kondo exchange that induces spin-flip inelastic tunneling between tip and sample to second order in perturbation theory\cite{Fernandez09,delgado10,delgado13}\cite{Fernandez09}. 
In this approach, the $dI/dV$ depends both on the occupations $P_M$ of the eigenstates of the spin Hamiltonian, $|M\rangle$, and
on the spin matrix elements $\langle M|S_a(i)|M^{'}\rangle$ where $S_a(i)$ is the spin operator or the spin site $i$ being excited with the STM tip and $a=x,y,z$.
We compute the occupations $P_M$ in two different approximations. First,  equilibrium, valid at low current, where $P_M$ is described by the Boltzman function.  At low temperatures, the only occupied state is the ground state, so that IETS is probing transitions from the ground-state only.  This gives rise to thermally broadened step-like $dI/dV$ curves. This approximation is justified as long as the  spin relaxation time of the on-surface spin states is much shorter than the time elapsed between inelastic tunnel events, controlled by the current intensity, that in turn can be controlled with the tip-surface distance.

Here we are interested in probing transitions from excited states too, so that we have to consider the high-current regime, where the on-surface spins do not relax before the next inelastic excitation event takes place.  In that regime, occupations are bias dependent and, more important, excitations from excited states can also be seen in the $dI/dV$, as shown experimentally for Mn dimers on Cu$_2$N \cite{loth10}.   
The non-equilibrium kinetics of the the occupations of the collective spin states of the molecules is governed by the Pauli master equation \cite{delgado10}:
\begin{equation}
    \frac{dP_M}{dt} = \sum_{M} P_{M'} W_{M',M} - P_{M} \sum_{M'} W_{M,M'}
    \label{eq:Pauli}
\end{equation}   
where $W_{M',M}$ are the transition rates from $M'$ to $M$.  Here we assume that these rates are given by the Kondo exchange interactions, including both the events where the electrons scatter between tip and sample and events where the electrons scatter between states in the same electrode. Expressions for $W_{M,M'}$ in terms of the spin matrix elements, are given in the appendix .  Equation (\ref{eq:Pauli}) is solved numerically using Runge-Kutta method.  The resulting $dI/dV$ lineshapes depart from the thermally broadened steps, on account of the bias dependence of the occupations, and result in overshoots at the inelastic step transitions. More importantly, the non-equilibrium $dI/dV$ feature inelastic steps associated to transitions between excited states. These features provide additional information about the energy levels, and thereby the Hamiltonian parameters.

\section{Characterization of the spin couplings by  IETS}

In this section we discuss how to determine the three exchange couplings of the BLBQBC interaction in a non-equilibrium IETS  experiment on a dimer. This would allow to determine how close a given system is of realizing the AKLT Hamiltonian.  Before we present the method with some detail, we   discuss the energy spectrum of the BLBQBC dimer.

\subsection{Phase diagram of the BLBQBC dimer}
Expressing the  $\vec{S}_1\cdot\vec{S}_2$ operator in terms of the total spin operator, it is straightforward to obtain the analytical expression for
the energy levels of the BLBQBC dimer:
\begin{align}
     E(S_T) &=\frac{J_1}{2} 
     {\cal F}(S_T)
     %\left[ S_T (S_T +1) - \frac{15}{2}\right] 
     + \frac{J_2}{4}  {\cal F}(S_T)^2 %\left[ S_T (S_T +1) - \frac{15}{2}\right]^2 \notag \\
     +  \frac{J_3}{8}{\cal F}(S_T)^3. 
     \label{eq:analytical_energies}
\end{align}
where ${\cal F}(S_T)=\left[ S_T (S_T +1) - \frac{15}{2}\right] $, and $S_T=0,1,2,3$ are the eigenvalues that the total spin operator of two spins $S=3/2$ can take. Thus, the dimer spectrum features, at zero magnetic field, four different multiplets, labelled with $S_T$.

\begin{figure}[t]
    \centering
    \includegraphics[width=1\linewidth]{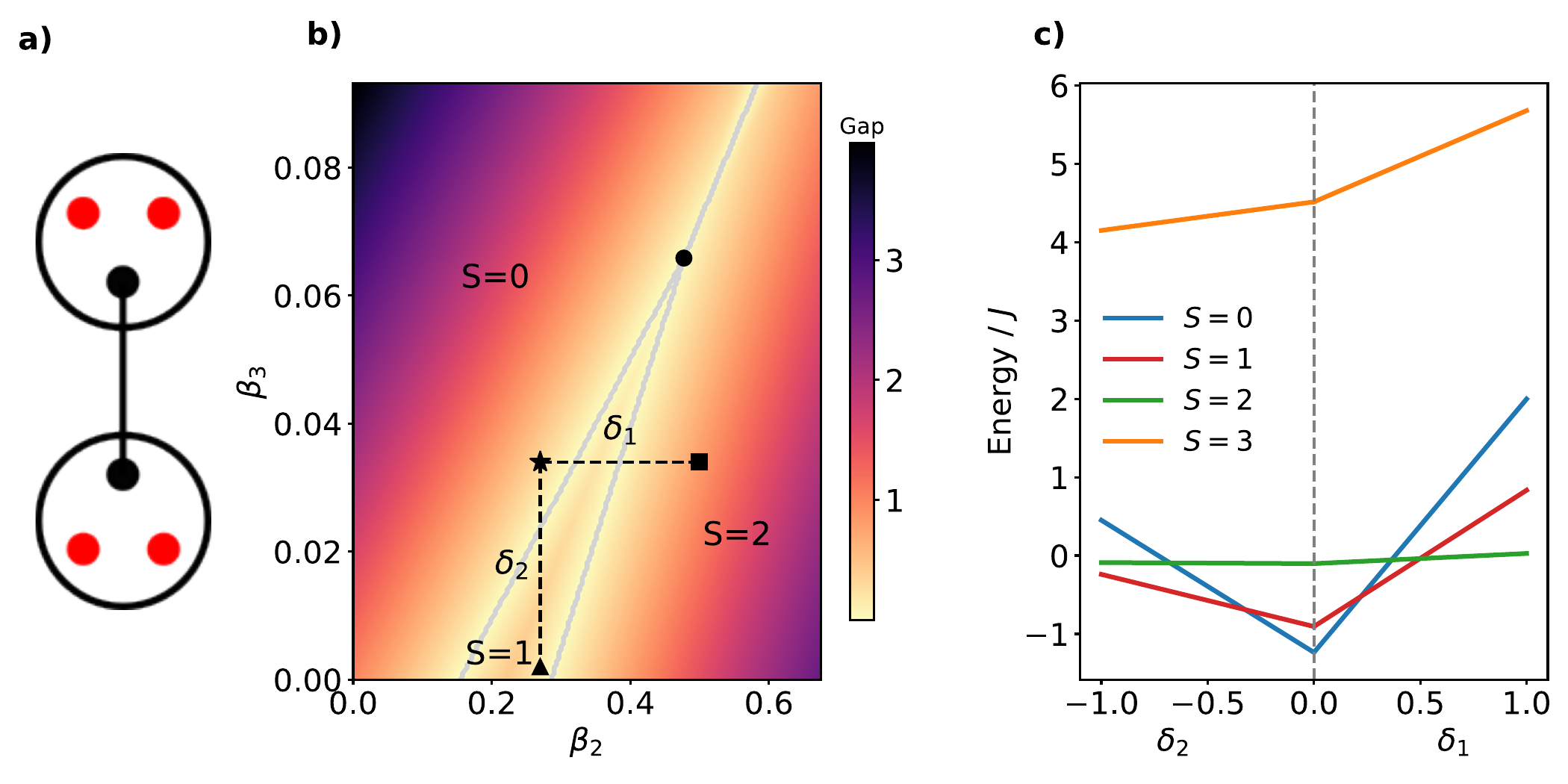}
    \caption{ a) Scheme of BLBQBC dimer, where the red dots represent the dangling virtual spins.  b) Phase diagram for the BLBQBC dimer , where the color map represents the magnitude of the gap as a function of $\beta_2,\beta_3$. We can find three regions, where the spin $S$ of the ground state is $S=0, 1, 2$. These three regions intersect  in the AKLT point. c) Energies of the multiplets going through paths $\delta_1$ and $\delta_2$. Here we can see how the energy changes when exploring the three different regimes of the phase diagram.  }
    \label{fig:dimerPD}
\end{figure}

Interestingly the spin of the ground state can take three different values, $S=0,1,2$ in the region where both $\beta_2,\beta_3$ are positive, and first neighbour exchange is antiferromagnetic.
In the Heisenberg point $(\beta_2=\beta_3=0)$ the spin of the ground state is of course $S=0$, on account of the antiferromagnetic nature of exchange. In contrast, in the AKLT point, the ground state has a nine-fold degeneracy containing spins $S=0,1,2$; this can also be interpreted based on the emergence of an effective $S=1$ at each site, resulting from the ferromagnetic coupling of pairs of dangling virtual spin-$1/2$.  In Fig. \ref{fig:dimerPD} we show the gap between the ground state and first excited state of the spin models as a function of $\beta_2,\beta_3$. This diagram defines three regions with different ground state spin; these three regions meet at the AKLT point. Marked with a star is the point in the phase diagram where a dimer made out of two nitrogen doped [5]-triangulenes (Aza[5]-triangulene) falls. Details on how these parameters were obtained for this particular system are given in Appendix \ref{app: aza5}. Interestingly, we see that for this system $\beta_2$ is large enough as to have $S=1$ as the ground state, but the size of $\beta_3$ takes it out of that region.  Hence, it is not unlikely that similar nanographene systems will be found where the dimer ground state   will be different from $S=0$. 

\subsection{Finding exchange values using IETS}

\begin{figure}[t]
    \centering
    \includegraphics[width =\linewidth]{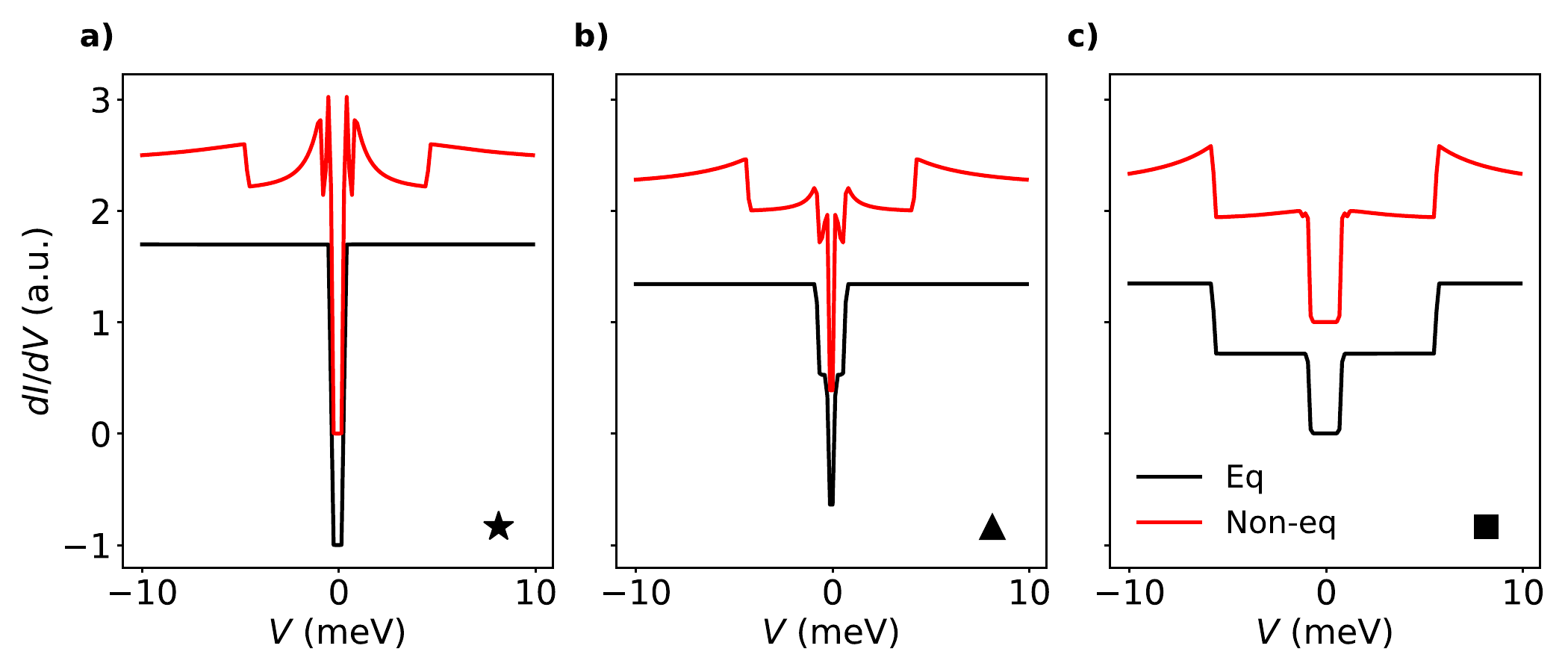}
    \caption{$dI/dV$ curves both for equilibrium IETS and non-equilibrium. The values used for the non-linear couplings of the BLBQBC model correspond to a) the aza-triangulene case with $\beta_2 = 0.27$ and $\beta_3=0.034$ ($\bigstar$ in Fig. 2), b) $\beta_2=0.27$ and $\beta_3=0.002$ ($\blacktriangle$) and c) $\beta_2=0.5$ and $\beta_3=0.034$ ($\blacksquare$). The parameters used for the simulations are shown in Appendix \ref{app:a}. }
    \label{fig:noneq_IETS}
\end{figure}

As we have seen before, the BLBQBC dimer has 4 multiplets ($S=0,1,2,3$), and thus three energy gaps. Here we will see how the parameters of the spin model can be experimentally obtained using IETS. 
First, in Fig. \ref{fig:noneq_IETS} we show in black the $dI/dV$ curves as a function of the applied bias using perturbation theory up to 2nd order, in the three regions of the $H_{\textrm{BLBQBC}}$ phase diagram. In panel a), where the ground state is a singlet, we find a single excitation step, associated to a singlet to a triplet transition. In panel b), because the ground state has $S=1$, we have an excitation step associated with the triplet-to-singlet transition, and then another step due to excitation of the $S=2$ manifold. Finally, in panel c) where the ground state is $S=2$, we find again two steps stemming from $S=2$ to $S=1$ and to $S=3$ transitions. %\redmark{I guess this sentence should be removed} In Appendix \ref{app:dIdV} we show the same plots, but this time accounting for 3rd order corrections, which give rise to a Kondo peak at zero bias when the ground state has $S>0$. 
From these $dI/dV$ plots, we see that we have at most two excitation steps. However, since the spin model we are studying has three parameters, we would need three inelastic steps to be able to determine the three energy scales
of the BLBQBC dimer  experimentally.

With this in mind, we move on from the equilibrium $dI/dV$, and compute its non-equilibrium counterpart. In this case, the occupation of the states is driven out of equilibrium by, for example, increasing the conductance of the junction bringing the tip closer to the surface\cite{loth10}. This leads to more available spin excitations, and it might become possible to probe enough steps to fully characterize the spin model. The observation of these three steps, at voltages $V_{I}$, $V_{II}$, $V_{III}$ (see red lines in  figure \ref{fig:noneq_IETS})  would permit one to fully determine a system of three equations and three unknowns ($J_1,J_2,J_3)$:
\begin{eqnarray} \label{eq:energy_transitions}
eV_{I}=E(1)-E(0)= J_1 - \frac{13}{2} J_2 + \frac{511}{16} J_3   
\nonumber \\
eV_{II}=E(2)-E(1)= 2 J_1 - 7 J_2 + \frac{163}{8} J_3 \nonumber \\
eV_{III}=E(3)-E(2)= 3 J_1 + \frac{9}{2} J_2 + \frac{189}{16} J_3 
\end{eqnarray}
where we have used ${\cal F}(0)=-\frac{15}{2}$, ${\cal F}(1)=-\frac{11}{2}$, ${\cal F}(2)=-\frac{3}{2}$ and ${\cal F}(3)=\frac{9}{2}$.   We note that the excitations $V_{II}$ and $V_{III}$ are, very often, very close in energy. Therefore,  in some cases, such as figure 3c, the two steps associated to them are almost degenerate, which will  complicate this procedure. 

\section{ IETS of  $S=3/2$ hexamer }
In this section, we examine the $S = 3/2$ AKLT model in a small cluster of six spins forming a hexamer. This structure is among the smallest that allows for geometric intuition while remaining computationally feasible via exact diagonalization (ED). Our focus is on understanding the role of in-gap edge states within the AKLT phase. To this end, we consider two geometries, namely: (i) \textit{open boundary conditions} (OBC), where the hexamer has an unpaired $S = 1/2$ edge spin at each site (Fig.~\ref{fig:Fig4dIdV}a), and (ii) \textit{periodic boundary conditions} (PBC), where the structure is "closed", passivating all edge spins and eliminating edge states.

\subsection{Low energy states at the AKLT point}

The AKLT wave function is built  by describing each physical $S=3/2$ with three virtual $S=1/2$, each forming a singlet with a virtual spin from a first neighbor.
 For the hexamer,  all 6 sites have coordination two, so that all of them have  one dangling virtual spin $1/2$ (red dots in Fig. \ref{fig:Fig4dIdV}a). This results in a ground state manifold with  degeneracy  $2^6=64$. We can also choose to introduce additional couplings among different pairs of spins in the hexamer,   reducing thereby both the number of doubly-coordinated sites and  the degeneracy of the ground state. For the case with periodic boundary conditions,  with all edges passivated,  the ground state is unique and has $S=0$, and the lowest excited state has $S=1$, with an excitation energy $\Delta E_{PBC} \simeq J/2$. The spectrum can be seen in the left side of Fig. \ref{fig:Fig4dIdV}c %\redmark{Are you sure the figure is given in units of $J$? The gap does not seem no be 0.5J from the figure}.

In an hypothetical finite-size nanographene hexamer that realizes the AKLT Hamiltonian the 64 degenerate ground states  include one septet $(S=3)$, five quintets $(S=2)$, nine triplets $(S=1)$, and five singlets $(S=0)$. The first excited state is now defined as the transition from this degenerate manifold to the 65th state, and has a value of $\Delta E_{OBC} \simeq 1.855 J$. This excitation energy is considerably larger than the PBC case. Intriguingly, we note that the lowest-energy excited state of the OBC hexamer, i.e. state number 65, has $S=4$.  For PBC the first excited state has $S=1$. We interpret the $S=4$ state with OBC as a collective excitation with $S=1$ over a ground state with $S=3$, as if the  collective excitation was mediating a ferromagnetic interaction among the dangling spins.

\subsection{IETS in the AKLT limit}

\begin{figure}[t!] 
    \centering
    \includegraphics[width=1\linewidth]{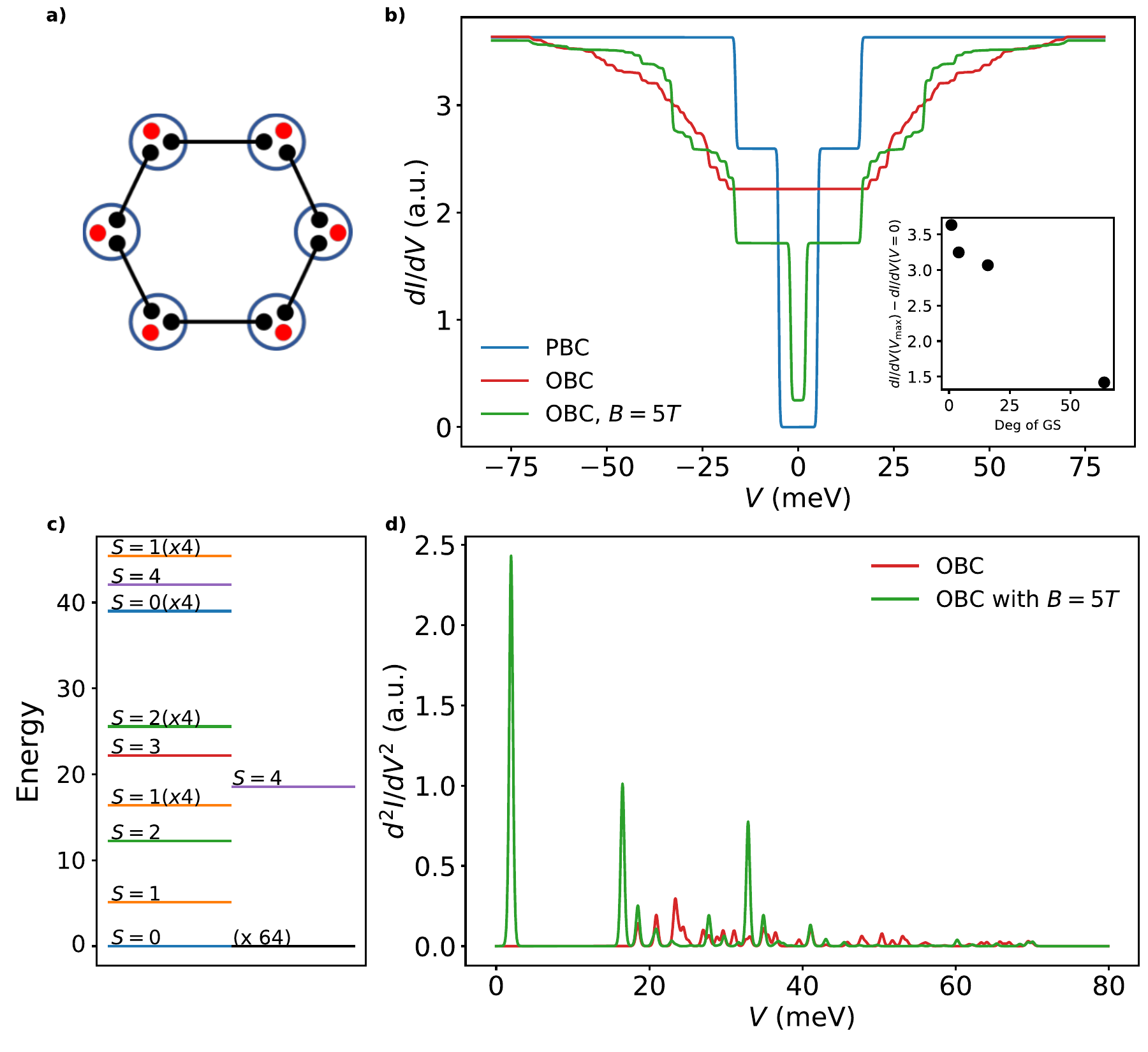}
    \caption{ a) Schematic representation of the AKLT hexamer. b) $dI/dV$ curves. Inset: Difference between the high bias and zero bias value for different number of passivated edge states. c) Energy spectrum of the hexamer in the AKLT point for PBC (left) and OBC (right). d) $d^2I/dV^2$ of the OBC hexamer with and without magnetic field. For all results we have used $J=10 meV$.  }
    \label{fig:Fig4dIdV}
\end{figure}

The inelastic electron tunneling spectroscopy (IETS) of a hexamer at the AKLT point is primarily dictated by the degeneracy of its ground state. With PBC there is a unique ground state and therefore transitions are only possible to excited states with $S=0$ and $S=1$.  Our calculations show two dominant steps in that case.  In contrast, with OBC, the ground state includes states with  $S = 0, 1, 2, 3$, which increases dramatically the number of possible spin excitations, leading to a complex $dI/dV$ spectrum with multiple inelastic steps associated (Fig.\ref{fig:Fig4dIdV}b).  

A crucial feature of the IETS spectrum of the hexamer in the AKLT point is that it shows a strong response  to an external magnetic field. When a magnetic field is applied, the ground-state degeneracy is lifted due to Zeeman splitting. In the limit where $g \mu_B B \gg k_B T$, states with larger negative $S_z$ become preferentially occupied. This has three main effects in the $dI/dV$ spectrum: i) steps associated to transitions between the previously degenerate 64-state manifold appear; ii) the number of transitions to higher energy states is reducing significantly, since the ground state is now unique (with a given $S$ and $S_z$ quantum numbers), and spin selection rules only allow transitions with $\Delta S = 0,1$; iii) the value of the conductance at zero bias increases. In fact, in the inset of Fig.\ref{fig:Fig4dIdV}b we find an approximate linear dependence between the degeneracy of the ground state and the conductance value at $V=0$.

The difference between  of the $B=0$ and large $B$ IETS spectra is better appreciated by representing $d^2I/dV^2$ (Fig.\ref{fig:Fig4dIdV}d).  Therefore, a smoking gun of the realization of the AKLT phase would be a very strong magnetic field dependence of the number of peaks in the  IETS spectra in finite-size lattices, revealing the topological degeneracy of the ground state.  

%On the other hand, when all edge states are passivated—such that there are no unpaired dangling spins—the system's ground state becomes unique with total spin $S = 0$. In this case, the complexity of the IETS spectrum is significantly reduced. Since there are no low-energy spin degrees of freedom available for inelastic transitions within the ground-state manifold, the number of observed inelastic steps in the $dI/dV$ curve is minimal. The primary excitation visible in the spectrum corresponds to a single transition from the singlet ground state to the first excited state with $S = 1$, occurring at an energy gap of $\Delta E_{PBC} \simeq J/2$. This result is in stark contrast to the case before with unpassivated edges, where a much richer spectrum emerges due to the interplay of multiple low-energy spin configurations. There results can be seen in Fig. \ref{fig:Fig4dIdV} b.
%\redmark{I think that either all energies are given in units of $J$ or none is in Fig4}

%\subsection{IETS OBC AKLT Point, with magnetic field}

%Discussion of IETS: OBC vs PBC: Many excitations, on account
%of having 64 ground states. 

%Discuss results at finite field: "dramatic filtering" of the IETS, as only the $S=3,S_z=3$ state becomes occupied. Transitions to $S=2,3,4$ manifolds above the gap, and $S_z=2$ states in the GS manifold are possible. 

%\begin{itemize}
%    \item   Two type of transitions:1)  "in-gap", between "edge" states    2) off-gap . Several excited states of this second type.   
%    Excitation of the $S=4$ state ("polaron?").  
%    \item  Mention This type of dichotomy should apply as well to larger systems 
%\end{itemize}

%When adding a magnetic field, the multiplets split according to their $S_z$. In the PBC case, this splitting will just show three steps instead of one, corresponding to the splitting of the triplet of the $1$st excited state with S=1. Instead, for OBC, things get more interesting. The degenerate states will also split according to their $S_z$ (Fig. \ref{fig:Fig3} d). The ground state will be now the state with more negative $S_z$, in this case with $S_z=-3$. Now, when using IETS, two different type of transitions can be observed. The first and more intense transition that will be visible is "in-gap", meaning that it will be in between this block of 64 degenerated states. The only allowed transition will be to the states with $S_z=-2$. The other type of transitions will be "off-gap". The first one will be of course to this $1$st excited state with $S_z=-4$, but other excited states will also show a step in the $dI/dV$ curve. This different types of transitions should be present also when going to larger systems. 

\subsection{Effective spin model for fractional edge spins  close to the AKLT point }
\begin{figure}[t]
    \centering
    \includegraphics[width=1\linewidth]{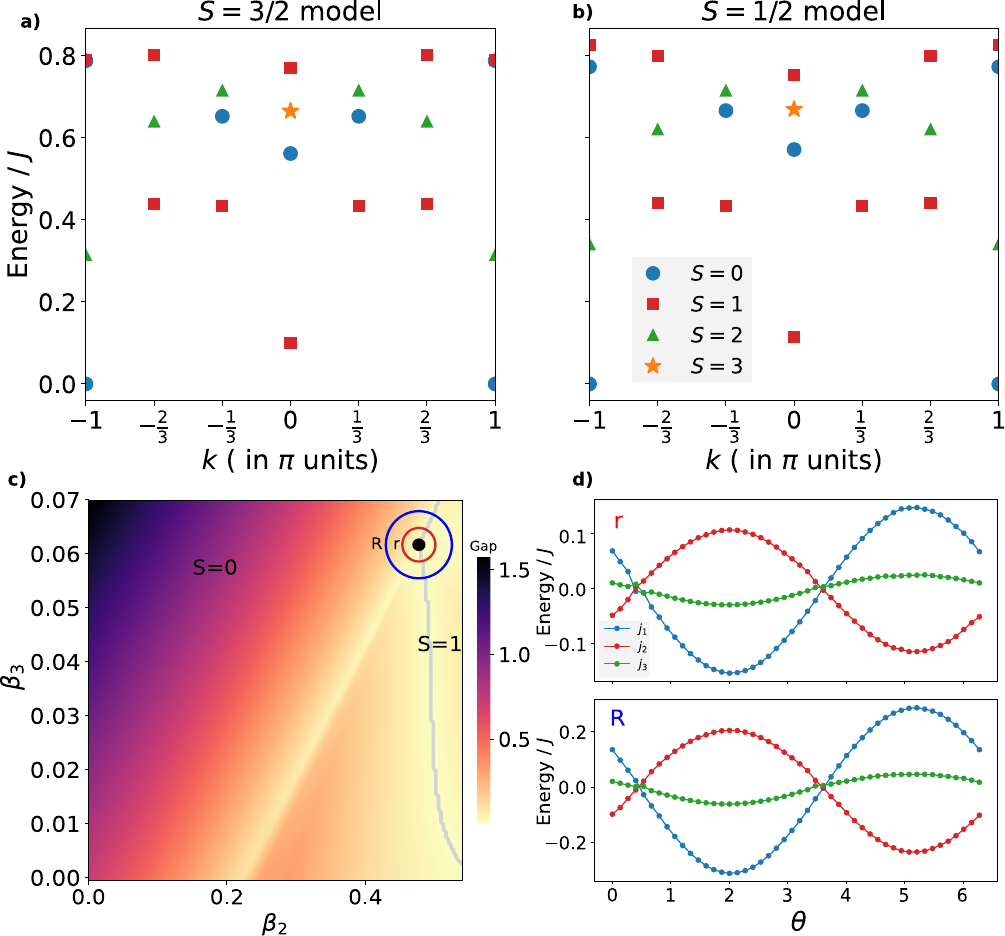}
    \caption{Effective model. a) and b) Comparison of the spectra of the $S=\frac{3}{2}$ model with the effective $S=\frac{1}{2}$ model. The multiplets have been labeled by their $S_{\rm tot}$ and in the $x$-axis we show their $k$ value. c) Value of the gap for different values of $\beta_2$ and $\beta_3$ in the BLBQBC model. In comparison with the dimer case, here we can only get a ground state with $S=1$ but not with $S=2$. d) Fitting of the values of $j_1$, $j_2$ and $j_3$ for the effective model for values of $\beta_2$ and $\beta_3$ around the AKLT point in two different trajectories with $r=0.05$ and $R=0.1$.   }
    \label{fig:EffectiveModel}
\end{figure}

We now discuss the spectrum of the OBC hexamer described with the BLBQBC model close to the AKLT point, 
so that the degeneracy of the GS is lifted, but there is still a clear gap between the 64 low energy states and the rest.  
In Fig.~\ref{fig:EffectiveModel}a, we present the energy spectrum of the BLBQBC model with parameters $\beta_2 = 0.4602$ and $\beta_3 = 0.0681$, chosen to be close to the AKLT point and in the red trajectory $r$ shown in Fig.~\ref{fig:EffectiveModel}c. The eigenstates are classified according to their total spin $S$ and wave vector $k$ (see Appendix \ref{ap:C6}).
Very much like in the case of Haldane chains, here we interpret the splitting of
the GS levels in terms of an {\em effective spin interactions } between the $S=1/2$ fractional spins.
 We propose the following effective Hamiltonian, that includes linear exchange couplings up to third-nearest neighbors:
\begin{equation}%\label{eq:linearH}
    H_{L}=\sum_{n=1,2,3} j_n \sum_{i}\vec{\sigma}_{i}\cdot\vec{\sigma}_{i\pm n}.
   % +j_2 \sum_{i}\vec{\sigma}_{i}\cdot\vec{\sigma}_{i\pm 2}
   % + j_3 \sum_{i}\vec{\sigma}_{i}\cdot\vec{\sigma}_{i+3}
   \label{eq:heff}
\end{equation}
This Hamiltonian can be resolved analytically for the case with $j_2=j_3$\cite{kouzoudis98}. 
For the general case, we  find $j_1,j_2,j_3$ by  numerical fitting. Specifically, we minimize the function:
\begin{equation}
    {\cal E}=\sum_{S,k,\alpha} \left(E^{(3/2)}_{S,k}(\alpha)-E^{(1/2)}_{S,k}(\alpha,j_1,j_2,j_3)\right)^2
\end{equation}
where the sum over $S$ and $k$ includes the 20 multiplets of the subspace of dimension $2^6$ discussed in the previous section and labeled with the correspondent $k$ value, and $\alpha$ is the additional index to label different multiplets with the same $S$ and $k$. The resulting spectrum, for $\beta_2 = 0.4602$ and $\beta_3 = 0.0681$, is shown in Fig. \ref{fig:EffectiveModel}b. We see a perfect agreement between the spectra of the full $S=3/2$ model and the effective model. Both models predict a ground state with $S=0$ and $k=\pi$, a first excited state with $S=1$ and $k=0$ and a second excited state with $S=2$ and $k=\pi$; as for the multiplets with higher energy one finds a perfect correspondence with the $S=3/2$ model.    

%\redmark{Este párrafo lo quitamos entero, no?}
%We note that there are some discrepancies for higher energy excited states, which suggest that there are other non-linear exchange terms in the effective Hamiltonian. Therefore, Hamiltonian (\ref{eq:heff}) is to be considered as a good approximation to the true effective Hamiltonian. In order to test this statement, we have computed the low energy spectrum for hexamers where only two spins have coordination 2. This is done by passivating four of the edge spins, leaving only two dangling  spins. The process can be done so that the two unpassivated spins are either first neighbours, second neighbours and third neighbours. In all cases the ground state manifold features a singlet and a triplet, separated by an energy difference that we compare with the values of $j_i$ obtained by fitting.  By doing this, we obtain $j_1 = 0.035 J$, $j_2 = -0.03 J$, and $j_3 = 0.037 J$, \redmark{to be compared with ...., for $\beta_2=, \beta_3=$}. Therefore, we see how that the values of $j_1,j_2,j_3$ obtained for the case of complete OBC have a sensible physical meaning. 

Further validation of the model comes from the fitting of the parameters $j_1,j_2,j_3$ as $\beta_2$ and $\beta_3$ are varied around the AKLT points, in two circular trajectories in the $(\beta_2,\beta_3)$ plane, centered around the AKLT point, with radius $r= 0.05,R= 0.1$, parametrized by $\theta\in (0,2\pi)$. Our results show a smooth evolution of $j_i(\theta)$, both for $r$ and $R$. We note that the effective exchange interaction is smaller for the smallest radius $r$, as expected, given that the effective couplings should vanish at the AKLT point.   Importantly, the $j_i$ couplings can take both positive and negative values, describing antiferromagnetic (AF) and ferromagnetic (FM) interactions, respectively.

The excursion in the $(\beta_2,\beta_3)$ plane visits two regions with different spin $S$ of the ground state, $S=0$ and $S=1$. The sign of the values obtained for $j_1,j_2,j_3$ provides insight to understand these two ground states. There is a segment within $0.6 \lesssim \theta\lesssim 3.7$  where $j_1$ and $j_3$ are AF, whereas $j_2$ is FM, that clearly stabilizes Neel-type correlations, compatible with $S=0$. On the contrary, out of this segment the signs of the 3 couplings are inverted, which results in frustration, and roughly correspond to the region where the ground state has $S=1$. In the proximity of the AKLT point where we are studying the system, the 65th state still has $S=4$ and $k=\pi$, and the big gap with the bulk of low-energy states is also mantained.

\section{Conclusions}

We have developed a theoretical framework to guide experimental efforts in realizing the AKLT model using nanographene-based structures. By characterizing the exchange couplings and predicting observable features of the AKLT phase, we highlight the crucial role of emergent fractional edge spins  in understanding the system’s low-energy physics. For the BLBQBC dimer, we show that the spin of the ground state can take  different values (\( S = 0, 1, 2 \)) depending on the values of $\beta_2$ and $\beta_3$, demonstrating the system's tunability of the system, where small variations in interaction parameters can lead to qualitative changes in the ground state and, consequently, in the excitation spectrum. Additionally, we propose a method to extract nonlinear exchange parameters using non-equilibrium inelastic electron tunneling spectroscopy (IETS), enabling full characterization of the spin Hamiltonian through differential conductance measurements.  

Extending our analysis to an hexamer in the $S = 3/2$ AKLT model, we predict a strong dependence of the IETS spectrum on an external magnetic field, leading to ground-state  degeneracy lifting and therefore the appereance of inelastic transition within the low-energy manifold, and suppression of high-energy excitations due to preferential occupation of states with large negative $S_z$. Near the AKLT point, the 64-fold ground-state degeneracy is lifted and the system can be described by effective interactions between fractional $S=1/2$ edge spins, which we model using an effective spin Hamiltonian incorporating up to third-nearest-neighbor interactions. An intriguing feature of the hexamer's excitation spectrum is the $S = 4$ state (state 65), which persists as the system moves away from the AKLT point. This state can be interpreted as a collective excitation mediating an effective ferromagnetic interaction among the dangling edge spins, contrasting with the periodic boundary condition (PBC) case, where the first excited state has $S = 1$. Future studies should extend these methods to larger nanographene structures, leveraging advanced numerical techniques like Quantum Monte Carlo \cite{sandvik10} and Neural-Network Quantum States \cite{carleo2017} to further explore AKLT physics in realistic experimental settings, with implications for quantum magnetism and measurement-based quantum computing\cite{wei22}.

\vspace{1cm}

\section*{Acknowledgements}

JFR acknowledges Nils Krane for conversations in the early stage of this project.
MFC
 acknowledges funding from Generalitat Valenciana
(CIACIF/2021/434).
J.F.-R. and J.C.G.H. %and A.C 
acknowledge financial support from 
%1
 SNF Sinergia (Grant Pimag),
 FCT (Grant No. PTDC/FIS-MAC/2045/2021),
 %3
 the European Union (Grant FUNLAYERS
- 101079184).
J.F.-R. acknowledges funding from
%
% 4
Generalitat Valenciana (Prometeo2021/017
and MFA/2022/045)
%5
and
MICIN-Spain (Grants No. PID2019-109539GB-C41 and PRTR-C1y.I1) 

\appendix 

\section{Calculation of $dI/dV$} \label{app:a}
For the sake of completeness,  we review the  formalism to calculate $dI/dV$ that we have used,
that follows previous work by one of us \cite{Fernandez09,delgado10}. The starting point is a Hamiltonian for free fermions in the tip and the surface, and a spin Hamiltonian, the BLBQBC model in this paper, for the on-surface spins.  The otherwise free fermions are coupled to the on-surface spins via a  Kondo exchange that has an additional degree of freedom (tip/sample). Therefore, there are 4 Kondo couplings, attending two the initial and final electrode of a given process: tip-tip (TT), tip-sample (TS), sample-tip (ST), sample-sample (SS). Only ST and TS processes contribute to the spin-flip  tunnel current.  TT and SS process play a role in the dissipative spin dynamics of the on-surface spins, relevant for the non-equilibrium calculation.

The  inelastic current  can be computed as \cite{Fernandez09,delgado10} the sum over initial states $M$ of the product of their occupations $P_M$
times the sum over final states $M'$, weighted by the spin matrix elements 
\begin{equation}
    I \propto %\frac{gs}{G_}
    \sum_{M,M'}P_M\left(V\right) i_{-} \left( \Delta_{M,M'} + eV\right) \times \sum_{\eta, \eta'}|\mathbf{S_{a, \eta, \eta'}^{M,M'}}|^2 
    \label{eq:noneq_current}
\end{equation}
where $\eta$ and $\eta'$ are the electrodes labels, T for tip and S for surface, and 
\begin{equation}
    \mathbf{S_{a, \eta, \eta'}^{M,M'}}  \equiv \frac{1}{\chi}\sum_{i,a=x,y,z } \nu_{\eta} (i) \nu_{\eta'} (i)\langle M|S_a(i)|M^{'}\rangle.
\end{equation}
The    dimensionless factors $\nu_{\eta}(i)$ code the intensity of the coupling of tip and sample to the different sites $i$ in a given  spin lattice\cite{delgado10}. Therefore, the site dependence of the inelastic current is encoded in the $\nu$ parameters. We also define a normalization factor  $\chi = \sum_i \nu_T(i)\nu_S (i)$ is a parameter that quantifies the tip-surface transmission through the magnetic atoms.
%where 
%the spin matrix elements between eigenstates $M$ and $M'$ are given by: 
%
%\begin{equation}
%    |S_{a}(n)^{M,M'}|^2=  |\langle M|S_a(n)|M^{'}\rangle|^2
    %S_{a,\eta,\eta^{'}}^{M,M'} (n)
    %
    %\equiv  \frac{1}{\chi} \sum_i \nu_{\eta}(i) \nu_{\eta^{'}}(i) \langle M|S_a(i)|M^{'}\rangle.
%\end{equation}
%
%where $\nu_{\eta}(i)$ are dimensionless parameters that quantify the strength of the exchange interaction of the on-surface spin $i$ with electronic states in the tip $(\eta=T)$ or surface ($\eta=S$), 
%$\chi=\sum_i \nu_T(i) \nu_S (i)$ is normalization factor
%and $a=x,y,z$.  
%Therefore, the parameters $\eta$ contain the information about what spin is being excited by the tip .
The function $i_-$ is given by\cite{delgado10}:
\begin{equation}
    i_{-} \left(\Delta + eV\right) = \frac{G_0}{e} \left[ {\cal G}\left(\Delta + eV\right) - \left(\Delta - eV\right) \right]
\end{equation} 
and  $G_0= \frac{e^2}{h}$
with ${\cal G} \left(\omega\right) \equiv \omega \left( 1- e^{-\beta\omega}\right)^{-1}$.
Using Eq. (\ref{eq:noneq_current}),  the calculation of $dI/dV$ gives thermally broadened steps at the bias voltages that match on-surface excitations that conserve energy and spin. The intensity of the steps is controlled by the spin matrix elements.

%For the calculation of the current operator, only the off-diagonal matrix elements  
%$ S_{a,S,T}^{M,M'} $ and $ S_{a,T,S}^{M,M'} $ contribute. In contrast, 
%the $\eta=\eta'$ matrix elements are relevant for the dissipative on-surface spin kinetics \cite{delgado10}, described by a Pauli master equation\cite{delgado10}, that we solve to obtain the voltage dependent occupation  $P_M(V)$ of the state $M$. 

In our simulations for the non-equilibrium kinetics of the spin dimer, we  take into account Kondo exchange interactions up to second order in perturbation theory.   Kondo exchange includes not only the spin and the momentum of the itinerant electrons but also an index for the electrode. Therefore, for the calculation of the spin relaxation and excitation rates we consider\cite{delgado10} scattering events where the initial and final states of the electron can be in two electrodes. As a result, there are 4 types of event, tip-tip, surface-surface, surface-tip, and tip-surface. For a given sign of bias, only the tip-surface events contribute to the excitation of the spin towards higher energy states. The tip-tip and surface-surface scattering events contribute to the relaxation of the system, promoting the occupation of low energy states.  The scattering rates depend on how the two sites of the dimer are coupled to tip and substrate.  We assume that the Kondo coupling to the substrate is not the same for the two spins in order to have finite spin relaxation rates \footnote{This is a shortcoming of the approximation of ignoring the Bloch phase in the Kondo coupling \cite{delgado2017}}. 
The expression for the rates that enter the master equation reads:
\begin{equation}
    W_{M,M'}^{\eta, \eta'} = \frac{2\pi T^2}{\hbar}  \mathcal{G}(\Delta+\mu_{\eta}-\mu_{\eta'})\frac{\rho_{\eta}\rho_{\eta'}}{4} \sum_n |\mathbf{S_{a,\eta,\eta'}^{M,M'}} (n)|^2
\end{equation}

where $T$ gives the magnitude of the exchange coupling of the atomic spin to the transport electrons, $\mu_{\eta}-\mu_{\eta'}$ is the voltage difference of the electrodes and $\rho_{\eta}$ is the density of states at the Fermi Energy for the electrode $\eta$. 
  
% To do list for MAR:
%1) Copy equation for W from my paper with Fernando 
% 2) Copy equation 20 

  For the simulation, we took the following parameters: temperature $T=0.05 K$, magnetic field $B=0$, the dimensionless coupling to tip of $\nu_S(1)=\nu_S(2) = 1.5$, $\nu_T(1) = 2.5$, $\nu_T(2) =1.5$ and $\rho_T=\rho_S=0.1$. 
  
%\redmark{I would show the parameters in the figures where they were used}

\section{Spin model parameters for the Aza[5]-Triangulene dimer} \label{app: aza5}

In the main text, we considered as a reference value for the $H_{\textrm{BLBQBC}}$ parameters the ones of an Aza[5]-Triangulene dimer, $\beta_2 = 0.27$ and $\beta_3 = 0.034$. In this appendix, we shall briefly explain how these values were obtained.

The Aza[5]-Triangulene (A5T) refers to a triangulene with five benzene rings along its side, where the central carbon atom was replaced by a nitrogen one \cite{anindya2022controlling,pawlak2025surface,villasvarela2023surface,wang2022aza,lawrence2023topological,yu2024prediction}. To model this molecule, we use a Hubbard model with first and third neighbor hopping, and an on-site potential on the nitrogen site and its nearest neighbors. A similar procedure has been used in Refs. \cite{catarina2023,henriques2024beyond}, showing results in excellent agreement with density functional theory (DFT). Based on those works we consider the first and third neighbor hoppings to be $t=-2.7$eV and $t_3 = t/10$, respectively, the Nitrogen on-site potential to be $V_0=-4$ eV (and $V_1 = -0.85$ eV on the first neighboring sites) and we take the Hubbard repulsion parameter as $U=|t|$. Without nitrogen functionalization, in the non-interacting limit, the 5T molecule has 4 orbitals at zero energy\cite{fernandez07}, each hosting one electron, leading to a spin $S=2$ when interactions are accounted for. However, when the central carbon atom is replaced by nitrogen, one of these zero energy levels goes down in energy, due to the electrostatic potential created by the extra proton in nitrogen, compared to carbon. This red-shifted zero mode becomes doubly occupied due to the additional electron brought in by the dopant; the other three levels at zero energy remain singly occupied, so that the ground state of  A5T  has $S=3/2$. The A5T  dimer is simply obtained by covalent coupling two of these systems tip to tip. Our calculations for the dimer predict   a $S=0$ ground state, with a set of low energy excitations well described by the spin-$3/2$ $H_{\textrm{BLBQBC}}$ Hamiltonian.

To obtain the values of the exchange interactions, we use a similar procedure to the one used in Refs. \cite{catarina2023,henriques2024beyond}. First, we solve the fermionic Hubbard model in the configuration interaction approach using the complete active space approximation (CI-CAS). Afterwards, we fit the energies of the spin model to the energies found from the fermionic model, allowing us to find the values of exchange one should use. The CI-CAS method can me summarized as follows: first, one solves the single particle problem (i.e. without Hubbard repulsion); second, the Hubbard Hamiltonian is expressed in the basis of the eigenstates found in the previous step; finally, we truncate the Hilbert space to include only the single particle states closer to zero-energy, and diagonalize the resulting Hamiltonian in the Hilbert space spanned by all possible electron configurations over the considered single particle states.

\section{Classification of hexamer states using $C_6$ symmetry} \label{ap:C6}

We discuss the method to label the states of hexamers with a wave vector, taking advantage
of the fact that hexamer Hamiltonians have $C_6$ symmetry.  We adopt the following procedure.
First, we build the translation operator, in terms of swap operators:
\begin{eqnarray}
    C_6^{S} &=& \prod_{i=0}^{4} {\cal S}^{(S)}_{i,i+1} .
\end{eqnarray}
where ${\cal S}^{(S)}_{i,i+1}$ is the swap operator for spin $S$. 
This operator satisfies the equation 
$C_6^{S} \ket{k} = e^{ik} \ket{k}$, where $k = \pi n /6$, with  $n = 0,\pm1,\pm2,3$. Generalization for $L$ sites is straightforward,
the upper limit on the product should be $L-2$ instead, and $k$ should take the values $\pi n/L$ with $n = 0, 1, ..., L-1$.

For $S=1/2$ the swap operators is:
\begin{eqnarray}
    {\cal S}^{(1/2)}_{ij}=  \left( \frac{1}{2} + 2 \vec{S}_i \cdot \vec{S}_{i+1} \right)
    \nonumber
\end{eqnarray}
For $S=3/2$, the swap operator adds quadratic and cubic terms and reads:
\begin{eqnarray}
    {\cal S}^{(3/2)}_{ij}=  -\frac{67}{32}- \frac{9}{8}(S_i \cdot S_{j}) + \frac{11}{18}(S_i \cdot S_{j})^2 
    %\notag \\
    + \frac{2}{9}(S_i \cdot S_{j})^3 
    \nonumber
\end{eqnarray}

The numerical diagonalizations of the hexamer Hamiltonians yield  manifolds with 
degeneracies larger than those imposed by the spin symmetry. 
We represent the translation operator in the basis of such manifolds and diagonalize it, obtaining thereby simultaneous eigenvectors of the Hamiltonian and the translation operator, with well defined wave vector $k$.

%%%%%%%%%%%

%\bibliographystyle{apsrev4-1}
\bibliography{bibshort}
\end{document}